\documentstyle[12pt]{article}

\title{
\begin{flushright}
{\normalsize Yaroslavl State University\\
             Preprint YARU-HE-94/01\\
             hep-ph/9404288} \\[3cm]
\end{flushright}
Double radiative decay $Z \rightarrow f\bar{f}\gamma\gamma$
       in the Standard Model}

\author{N.V.Mikheev\thanks{E-mail: \quad physteo@univ.yars.free.net}
       and A.Ya.Parkhomenko\thanks{E-mail:
                                   \quad uni@cnit.yaroslavl.su} \\
       {\small\it Division of Theoretical Physics, Department of
                                                      Physics,} \\
       {\small\it Yaroslavl State University, Sovietskaya 14,
                                             150000 Yaroslavl,} \\
       {\small\it Russia}}

\date{11 January 1994}

\begin{document}

\maketitle

\begin{abstract}

Z-boson decay $Z \rightarrow f\tilde{f}\gamma\gamma$ in the
Standard Model is analysed. The distribution function on the
invariant masses of the photon and fermion pairs is calculated
in the leading logarithmic approximation.
It is shown that this distribution function has a specific
shape of a ``crest''.

\end{abstract}

\newpage

\maketitle

Intensive investigations of $Z$-boson physics are carried out in
a process of $e^+ e^-$ annihilation at the LEP. In particular,
the processes of $e^{+}e^{-} \rightarrow f\bar{f}\gamma\gamma$
type (where $f$ is an arbitrary fermion) are analysed in detail.
For example, four such events with the lepton pair and the hard
photon pair (with the invariant mass $m_{\gamma\gamma} \simeq
60 \; GeV$) in the final state has been observed in the LEP data
collected in 1991 and 1992~\cite{L3l}. In these experiments the
center-of-mass energy of the $e^+ e^-$ pair is close to the
$Z$-boson mass, so the final state is mainly
the result of a real
$Z$-boson production and its decay into $f\bar{f}\gamma\gamma$.
For the ``hard'' photons ($m_{\gamma \gamma} > 20 \, GeV$)
a possible contribution to the processes
$e^+ e^- \rightarrow f \bar f \gamma \gamma$ with
the photon bremsstrahlung from the initial state is suppressed.
Only corrections connecting with the ``soft'' photon radiation
from the initial state are essential.
In the papers~\cite{Ja,S,JaW}
such type events either with photon bremsstrahlung from the initial
state or without it were estimated by numerical methods.
In the framework of the Standard Model the additional channel
$e^+ e^- \rightarrow Z H \rightarrow f \bar f \gamma \gamma$
with the Standard Higgs boson in the intermediate state is possible,
but its contribution to the full cross section
of this process is rather small.
For that reason the four events announced by the L3
collaboration~\cite{L3l} can not be explained by the manifestation
of the Standard Higgs boson with the mass $m_H \simeq 60 \, GeV$
(see, for example,~\cite{KK}).

In this paper we calculated analytically the $Z$-boson decay
$Z \rightarrow f\bar{f}\gamma\gamma$ in the framework of the
Standard Model in the approximation of leading logarithms.
We obtained a simple expression for the distribution function
on the invariant masses of the photon and fermion pairs.
It is shown that the distribution function has a shape
of a ``crest''.

The process $Z \rightarrow f\bar{f}\gamma\gamma$ in the Standard
Model is described by the set of six graphs (see Fig.1) and has
the following matrix element:

\begin{equation}
{\cal M} = - {e^{3}Q^{2} \over \sin 2\vartheta_{W}} \,
(J \varepsilon^{(Z)}),             \label{eq:M}
\end{equation}

\noindent where $eQ$ is the fermion charge,
$\varepsilon^{(Z)}_{\alpha}$ is the $Z$-boson polarization vector
and $J_{\alpha}$ is the four-current of the final state
$f\bar{f}\gamma\gamma$:

\begin{equation}
J_{\alpha} = \sum^{3}_{i=1} \left ( J^{(i)}_{\alpha} +
\tilde{J}^{(i)}_{\alpha} \right ).
\label{eq:J}
\end{equation}

\noindent The currents $J^{(i)}_{\alpha}$ and
$\tilde{J}^{(i)}_{\alpha}$ are calculated according to
the standard Feynman rules and have the following form:

\begin{eqnarray}
J^{(1)}_{\alpha} & = -(\bar{u}(p_{1}) \hat{\varepsilon}_{1}
(\hat{p}_{1} + \hat{k}_{1} - m_{f})^{-1} \gamma_{\alpha}
(g_{V} + g_{A} \gamma_{5}) (\hat{p}_{2} + \hat{k}_{2} + m_{f})^{-1}
\hat{\varepsilon}_{2} u(-p_{2})),  \nonumber \\
J^{(2)}_{\alpha} & = (\bar{u}(p_{1}) \hat{\varepsilon}_{1}
(\hat{p}_{1} + \hat{k}_{1} - m_{f})^{-1} \hat{\varepsilon}_{2}
(\hat{p}_{1} + \hat{k}_{1} + \hat{k}_{2} - m_{f})^{-1}
\gamma_{\alpha} (g_{V} + g_{A} \gamma_{5}) u(-p_{2})),
\label{eq:Ji} \\
J^{(3)}_{\alpha} & = (\bar{u}(p_{1}) \gamma_{\alpha} (g_{V} + g_{A}
\gamma_{5}) (\hat{p}_{2} + \hat{k}_{1} + \hat{k}_{2} + m_{f})^{-1}
\hat{\varepsilon}_{1} (\hat{p}_{2} + \hat{k}_{2} + m_{f})^{-1}
\hat{\varepsilon}_{2} u(-p_{2})), \nonumber
\end{eqnarray}
\begin{eqnarray}
\tilde{J}^{(i)}_{\alpha} & = & J^{(i)}_{\alpha} \bigg\vert_{
\mbox{\large $k_1 \leftrightarrow k_2 \atop
\varepsilon_1 \leftrightarrow \varepsilon_2$}} , \nonumber \\
g_{V} = T_{3} & - & 2 Q \sin^{2}\vartheta_{W},
\qquad g_{A} = T_{3}, \nonumber
\end{eqnarray}

\noindent where $m_{f}$ is the fermion mass,
$\varepsilon_{1}, k_{1}$ and $\varepsilon_{2}, k_{2}$ are the
polarization vectors and the four-momenta of the photons,
$p_{1}$ and $p_{2}$ are the fermion and antifermion four-momenta,
$T_{3}$~is the third component of the fermion weak isospin.

For further analysis it is useful to introduce the dimensionless
invariant masses of the photon pair $x$ and of the fermion pair $y$:

\begin{equation}
x^{2} = {(k_{1} + k_{2})^{2} \over m^{2}_{Z}} =
{m^{2}_{\gamma\gamma} \over m^{2}_{Z}} ,
\quad y^{2} = {(p_{1} + p_{2})^{2} \over m^{2}_{Z}}
= {m^{2}_{ff} \over m^{2}_{Z}}.      \label{eq:xy}
\end{equation}

\noindent The differential probability of the decay
$Z \rightarrow f\bar{f}\gamma\gamma$ with the hard photons
was calculated in the approximation of leading logarithms
$((p_{1}k_{1}) \ll m^{2}_{Z}, (p_{1}k_{2}) \ll  m^{2}_{Z},
(p_{2}k_{1}) \ll  m^{2}_{Z}, (p_{2}k_{2}) \ll  m^{2}_{Z})$.
The correspondent distribution function has a simple form:

\begin{eqnarray}
{1 \over N} {d^{2} \Gamma \over dx dy} & \equiv & F(x,y) \simeq
{y \over x} \; {I^{2}(x,y) + (1+y^{2})^{2} \over I^{2}(x,y) +
x y r_{f\gamma} \ln^{2}(r_{f\gamma})} \; I(x,y),
\label{eq:disf}  \\
I(x,y) & = & \left[ (1-x-y)(1-x+y)(1+x-y)(1+x+y) \right]^{1 \over 2},
\nonumber \\
N & = & 2 \left ( {\alpha \over \pi} \right )^2 \ln^2(r_{f\gamma})
\Gamma_f,       \nonumber \\
\Gamma_{f} & \equiv & \Gamma (Z \rightarrow f\bar{f}) =
{G_{F} \over \sqrt 2} {m^{3}_{Z} \over 6\pi }
\left [ g^{2}_{V} + g^{2}_{A} \right ],   \nonumber \\
r_{f\gamma} & = & {\Delta m^{2}_{f\gamma } \over m^{2}_{Z}} \ll  1 ,
\nonumber
\end{eqnarray}

\noindent where $m_Z$ is the $Z$-boson mass and
$\Delta m_{f\gamma}$ is the cut of the invariant mass
of the photon - fermion pair.
Let us note that in the leading logarithmic
approximation fermion mass effects may be negligible even for
heavy ($\tau$-lepton and $c$- and $b$-quarks) fermions.
The distribution function $F(x,y)$ ties the distributions in two
different regions on the $(x,y)$-plane:

\noindent i) the neighbourhood of the physical region
boundary $x+y=1$, where $I(x,y)$ is close to  zero  and
may be negligible in the ratio of the expression~(\ref{eq:disf});

\noindent ii) the region far from the boundary,
where the term containing $r_{f\gamma}$ in the
denominator of~(\ref{eq:disf}) may be negligible.

\noindent The integrating of the expression~(\ref{eq:disf})
over the invariant masses $x$ and $y$ with the cut of the photon
invariant mass gives the partial width $\Gamma(Z \rightarrow
f\bar{f}\gamma\gamma)$ of the $Z$-boson decay in the leading
logarithmic approximation:

\begin{eqnarray}
\Gamma & = & \left ( {\alpha \over \pi } \right )^2
\ln^2(r_{f\gamma }) \ln^2(r_{\gamma\gamma }) \Gamma_f,
\label{eq:G} \\
r_{\gamma\gamma} & = & {\Delta m^{2}_{\gamma\gamma}
\over m^{2}_{Z}} \ll  1,   \nonumber
\end{eqnarray}

\noindent where $\Delta m_{\gamma\gamma}$ is the cut
of the invariant mass of the photon pair.

The distribution function $F(x,y)$ with $\Delta m_{f\gamma} =
2 \, GeV$ (Fig.2) and with $\Delta m_{f\gamma} = 6 \, GeV$ (Fig.3)
is represented in the intervals $0.5 \le x \le 1$ and
$0 \le y \le 0.5$.
The distributions showing on Fig.2 and Fig.3  have a specific
shape of a "crest" which is parallel to the border of the physical
region $x+y=1$. The "crest" is dependent of the cut of the fermion -
photon invariant mass $\Delta m_{f\gamma}$: with decreasing mass cut
it becomes more narrow and steeper.
Events will be observed mainly in the neighbourhood of the
border $x+y=1$ either the statistics of the $Z \rightarrow
f\bar{f}\gamma\gamma$ events will be increased or the precision
of the angular resolution of the fermion - photon pair
will be improved.
Let us note that the parameters $x$ and $y$ of the four experimental
$\ell^{+}\ell^{-}\gamma\gamma$ events \cite{L3l} fall into the
examined region and in fact lie near the border of the physical
region (see Fig.4 of the preprint~\cite{L3l}).

The expression~(\ref{eq:disf}) allows to estimate the branch
of the "hard" radiative decay $Z \rightarrow f\bar{f}\gamma\gamma$
for different fermion flavours in the following way:

\begin{equation}
Br(x_{0}) = {N \over \Gamma_{0}} \int^{1}_{x_0} dx \int^{1-x}_{0}
dy F(x,y),           \label{eq:Br}
\end{equation}

\noindent where $\Gamma_{0} \equiv \Gamma (Z \rightarrow  all)$
is the total $Z$ - boson width  and $x_{0}$ is the
minimal value of the ``hard'' photon invariant mass
(we took $x_{0} = 0.5)$.
For the lepton pair production the expression (\ref{eq:Br})
gives $Br(0.5) = 2.8 \cdot 10^{-6}; 2.1 \cdot 10^{-6}$ and
$1.1 \cdot 10^{-6}$ for the mass cut $\Delta m_{\ell\gamma} =
2 \, GeV$; $3 \, GeV$ and $6 \, GeV$ correspondingly.
It does not contradict to the LEP statistic data on the leptonic
$Z$ - boson decays ($\sim 10^{6}$ events) \cite{L3l}:
the decays $Z \rightarrow \tau^{+}\tau^{-}\gamma\gamma$ with
$\Delta m_{\tau\gamma} \sim 6 \, GeV$ are absent and the number of
$\mu^{+}\mu^{-}\gamma\gamma$ events with $\Delta m_{\mu\gamma}
\sim 2 \, GeV$ (three  events) is larger than
$e^{+}e^{-}\gamma\gamma$ with $\Delta m_{e\gamma} \sim 3 \, GeV$
(one event). Near $Z$ resonance the processes with ``hard''
photons $e^{+}e^{-} \rightarrow \gamma\gamma$ + hadrons are mostly
the result of the quark-photon decays
$Z \rightarrow q\bar{q}\gamma\gamma$ of the $Z$-boson.
The quark decay probability differs from the leptonic one
by the factor:

\begin{displaymath}
C_{q} \simeq  12 Q^{4}_{q}
\left [ g^{2}_{V} + g^{2}_{A} \right ]_{q}.
\end{displaymath}

\noindent The factor $C_{q}$ for ``up'' and ``down'' quarks has
the values: $C_{u} \simeq 0.7$  and $C_{d} \simeq  0.15$.
Let us note that the value of the mass cut of the quarks forming
the hadronic jets is the same as or even greater than the one for
the $\tau$ - lepton.
Taking this into account we obtained from~(\ref{eq:Br})
the following limits for ``up'' and ``down'' quarks:
$Br_{u}(0.5) \le  0.9 \cdot 10^{-6}$ and $Br_{d}(0.5) \le  0.2
\cdot 10^{-6}$.
It is clear that on the recent level of the data
statistics~\cite{L3h} such processes are invisible.
In our approximation the processes $Z \rightarrow \nu\bar{\nu}
+ n \gamma \; (n=1,2,\ldots)$ are absent because the neutrino
has not the electric charge $Q_{\nu} = 0$.

In conclusion, we calculate the probability of the foundation of
$n$ events clustering within an arbitrary mass bin of 2 $\Delta x$
width inside the photon invariant mass region $x \ge x_{0}$ in the
following way:

\begin{eqnarray}
W_{n}(x_{0};\Delta x) & = & \sum^{L}_{i=1} \Big [ {P(x_{i} -
\Delta x \le x \le x_{i} + \Delta x) \over P(x_{0} \le x \le 1)}
\Big ]^{n} ,        \label{eq:clust}  \\
P(a \le x \le b) & = & \int^{b}_{a} dx \int^{1-x}_{0} dy F(x,y),
\nonumber
\end{eqnarray}

\noindent where $L = x_{0}/(2\Delta x)$ is the full number
of the 2$\Delta x$ width stripes in the region $x \ge x_{0}$
and $x_{i} = x_{0}+(2i-1) \Delta x \quad (i=1, \ldots, L)$
is the average value of $x$ in the stripe.
The estimation by~(\ref{eq:clust}) of the four clustering events
$(n=4)$ probability with the parameters $x_{0} \simeq 0.5$
and $\Delta x \simeq 0.025$ gives $W_{4}(0.5;0.025) = 1.5
\cdot 10^{-2}$.
It is not exclude the possibility of the describing of the four
events~\cite{L3l} within the Standard Model as the double radiative
decay $Z \rightarrow \ell^+\ell^-\gamma\gamma$.

\bigskip

{\large\bf Note added}

\bigskip

After the paper was completed, we received the preprint by
K.~Kolodziej, F.~Jeger\-leh\-ner and G.J.~van~Oldenborgh~\cite{KJO}
in which the process $e^+ e^- \rightarrow \mu^+ \mu^- \gamma
\gamma$ is analysed by numerical methods.
In particular, estimations of the cross section of this process,
expected number of such events in an experiment,
and the distribution function on the invariant masses of the
photon and muon pairs are represented.
The results of our paper agree with those of the preprint~\cite{KJO}.

\bigskip

{\large\bf Acknowledgment}

\bigskip

The authors are grateful to K.A.Ter-Martirosyan for permanent
interest and stimulating discussions, to A.D.Smirnov for fruitful
discussions and to A.V.Kuznetsov for helpful remarks.
This work was supported, in part, by a Soros Humanitarian
Foundations Grant awarded by the American Physical Society.

{\Large \bf Figure Captions}

\begin{enumerate}

\item Graphs describing the decay $Z \rightarrow f\bar{f}\gamma\gamma$.

\item The distribution function F(x,y) with the invariant mass cut
$\Delta m_{f\gamma}$ = 2 GeV.

\item The distribution function F(x,y) with the invariant mass cut
$\Delta m_{f\gamma}$ = 6 GeV.

\end{enumerate}

\end{document}